\documentclass[twocolumn,showpacs,preprintnumbers]{revtex4}
\usepackage{graphicx}
\usepackage{dcolumn}
\usepackage{bm}
\usepackage{amsmath}
\usepackage{hyperref}

\begin{document}

\title{Dynamics of Insulating Skyrmion under Temperature Gradient}
\author{Lingyao Kong$^{1}$}
\author{Jiadong Zang$^{1,2}$}
\email{jiadongzang@gmail.com} \affiliation{$^1$ State Key Laboratory
of Surface Physics and Department of Physics,
Fudan University, Shanghai 200443, China\\
$^2$ Department of Physics and Astronomy, Johns Hopkins University,
Baltimore, MD21218, U. S. A.}
\date{\today }

\begin{abstract}
We study the Skyrmion dynamics in thin films under a temperature
gradient. Our numerical simulations show that both a single and
multiple Skyrmions in crystal move towards the high temperature
region, which is the contrary to the particle diffusions. Noticing a
similar effect in the domain wall motion, we employ a theory based
on the magnon dynamics to explain this counter-intuitive phenomenon.
Different from the temperature driven domain wall motion, Skyrmion's
topological charge plays an important role, and a transverse
Skyrmion motion is observed. Our theory turns out to be in agreement
with the numerical simulations, both qualitatively and
quantitatively. Our calculation indicates a very promising Skyrmion
dynamic phenomenon to be observed in experiments.
\end{abstract}

\pacs{75.70.Kw, 75.10.Hk, 66.30.Lw, 75.40.Mg} \maketitle

A Skyrmion is a topological configuration in which local spins wrap
around the unit sphere for an integer number of
times\cite{skyrme1961}. After decades of theoretical
discussions\cite{Bogdanov, Rosler}, it has been recently observed in
the bulk sample of MnSi\cite{Muehlbauer2009a}. This material is a
typical helimagnet where inversion asymmetry induced
Dzyaloshinsky-Moriya (DM) interaction is significant; the latter
plays an important role to generate Skyrmion configurations. Their
neutron scattering study shows that Skyrmions perfectly pack
themselves in triangle crystals as a compromise between DM
interaction and ferromagnetic Heisenberg exchange. However, due to
the competition between conical phase, the Skyrmion phase
unfortunately survives only in a narrow window at finite temperatures\cite%
{Muehlbauer2009a}. Later, real space image in Fe$_{x}$Co$_{1-x}$Si thin film%
\cite{Yu2010a} has demonstrated that a Skyrmion crystal phase can be
considerably enlarged in two dimensions, and stable down to zero
temperature \cite{Yi2009, Rosler2010, Han2010}. Further exploration
shows that Skyrmion phases are not only present in these two
metallic materials, but also in
insulating materials like Cu$_{2}$OSeO$_{3}$\cite{Seki2012} and BaFe$%
_{1-x-0.05}$Sc$_{x}$Mg$_{0.05}$O$_{19}$\cite{Yu2012}.

After the discovery of Skyrmion crystals, numerous efforts has been
devoted to the manipulations of Skyrmions. Due to their topological
nature, Skyrmions remain stable against moderate perturbations.
Therefore, controlling the motion of Skyrmions would allow for
potential applications of Skyrmion physics. To this end, Skyrmion
dynamics has been discussed in detail\cite{Zang2011, Schulz2012,
Mochizuki2012}. One well accepted way to control the motion of
Skyrmions in metallic thin films is via a current. Unlike the
regular domain wall motion driven by the current, the Skyrmion
motion can occur at a tiny current threshold\cite{Jonietz2010}. This
advantage makes low-dissipative Skyrmion manipulation possible. An
interesting question is if it is possible to drive the motion of
insulating Skyrmions. If the answer is positive, one can thoroughly
get rid of the dissipations from the conducting current.

In this letter, we study the directional motion of insulating
Skyrmions under a temperature gradient. Insulating materials help us
to get rid of the influence from conduction
electrons\cite{Jonietz2010,Evershor2012}. Interestingly, our study
shows that Skyrmions unconventionally move towards high temperature
regions, contrary to the usual Brownian motion. A Skyrmion as a
large-size quasi-particle appears to have negative diffusive
coefficient. Followed by numerical simulations, a magnon assisted
theory is employed to explain this novel phenomenon.

%It is important to emphasize that we are concentrating on the insulating helimagnet in this letter. A previous experiment has been done on the temperature gradient on metallic Skyrmion crystal\cite{Jonietz2010}. However, that experiment is done on the metallic MnSi sample in the presence of a steady current. Therefore the intrinsic diffusion of Skyrmions is unseen. Insulating Skyrmion crystals provide ideal platform to study the diffusive properties of Skyrmions. Because of the absence of itinerating electrons, there would be no external transfer torque acting on Skyrmion.  However our study shows the diffusion of Skyrmion is unconventionally in the opposite direction.

\textit{Numerical Simulation:} To simulate the magnetization
dynamics at finite temperature, the stochastic
Landau-Lifshitz-Gilbert (LLG) approach is
employed\cite{Palacios1998,Hinzke}. The effect of the thermal
fluctuation at the temperature $T$ is characterized by a random
field $\mathbf{L}$ in addition to the usual LLG equation. The
equation of motion is given by
\begin{equation}
\mathbf{\dot{m}}=-\gamma \mathbf{m}\times (\mathbf{H}_{eff}+\mathbf{L}%
)+\alpha \mathbf{m}\times \mathbf{\dot{m}}  \label{eq_LLG}
\end{equation}%
where $\gamma =g/\hbar $ is the gyromagnetic ratio and $\alpha $ is
the Gilbert damping coefficient. The magnitude of the magnetization
$\mathbf{m}$ is normalized to unity. In the case of ferromagnetic
insulators, $\alpha $ can be tiny, due to the absence of conduction
electrons to dissipate the magnetization energy.
$\mathbf{H}_{eff}=-\partial H/\partial \mathbf{m}$ is the effective
field acting on the local magnetization $\mathbf{m}$. In order to
eventually achieve thermal equilibrium eventually, the
dissipation-fluctuation relation$\left\langle L_{\mu }(\mathbf{r},t)L_{\nu }(%
\mathbf{r}^{\prime },t^{\prime })\right\rangle =\xi \delta _{\mu \nu
}\delta (\mathbf{r}-\mathbf{r}^{\prime })\delta (t-t^{\prime })$ is
satisfied, where $\xi =\alpha a^{2}k_{B}T/\gamma $ and $a$ is the
lattice constant. The average $\left\langle {}\right\rangle $ is
taken over all the realizations of the fluctuation field. Here, a
uniform but small temperature gradient is assumed. As the thermal
fluctuation of each spin is about $k_{B}T$, which is about $J/10$ in
our simulation. As long as it is larger than the temperature
difference between neighboring sites, the local equilibrium can be
established and this stochastic LLG approach is justified. In this
case, $\xi $, together with $T$, is a linear function of the
position. In what follows, the temperature gradient is turned on
longitudinally along the $x$ direction. Numerically, the stochastic field $%
L_{\mu }(\mathbf{r},t)$ is created by a random number generator with
the mean square controlled by the temperature, and the stochastic
LLG equation
Eq.(\ref{eq_LLG}) is integrated out in the deterministic Heun scheme\cite%
{Palacios1998}, with a time step of $0.05\hbar /J$. The initial
Skyrmion configurations are given by classical Monte Carlo updates
followed by further relaxation realized by solving the LLG equation
with a fourth order Runge-Kutta method\cite{Yu2010a}.

%%%%%%%%%%%%%%%%%%%%%%%%%%%%%%%%%%%%%%%%%
%\begin{figure}
%\begin{center}
%\begin{tabular}{cc}
%$H=0.3$ & $H=0.15$\tabularnewline
%\includegraphics[width=4.5cm]{S3.eps} & \includegraphics[width=4.5cm]{X3.eps}\tabularnewline
%\end{tabular}
%\caption{Snapshots of Skyrmion motions.}
%\par\end{center}
%\label{fig_snapshots}
%\end{figure}
%%%%%%%%%%%%%%%%%%%%%%%%%%%%%%%%%%%%%%%%
%%%%%%%%%%%%%%%%%%%%%%%%%%%%%%%%%%%%%%%%%
\begin{figure}[tbp]
\centering
\includegraphics[width=8cm]{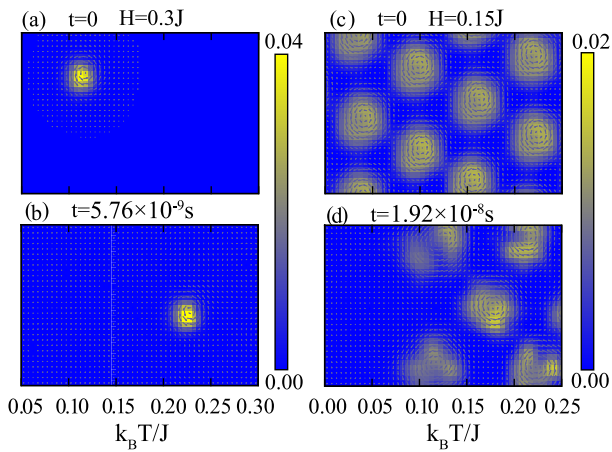}
\caption{Snapshots of Skyrmion motions. Color bars stand for the
topological charge density $q$. At the critical magnetic field of
$H=0.3J$, single Skyrmion is generated (a). Under temperature
gradient, it moves from low to high temperature (b). Skyrmion
crystal moves in a similar way under a lower magnetic field
$H=0.15J$ (c, d).} \label{fig_snapshots}
\end{figure}
We employed the standard model given by
\begin{equation}
H=\sum_{\langle ij\rangle }[-J\mathbf{m}_{i}\cdot \mathbf{m}_{j}+\mathbf{D}%
_{ij}\cdot (\mathbf{m}_{i}\times
\mathbf{m}_{j})]-\sum_{i}\mathbf{H}\cdot \mathbf{m}_{i}
\label{eq_discret Hamiltonian}
\end{equation}%
where the DM vector $\mathbf{D}_{ij}=D\hat{\mathbf{r}}_{ij}$ points
from one local magnetization to the other. The magnetic field
$\mathbf{H}=H\hat{z}$
is perpendicular to the film. $H$ relates the real magnetic field $h$ by $%
H=\mu _{B}h$. In the simulations, the Heisenberg exchange
$J/k_{B}=50K$, and the strength of DM interaction $D=0.5J$. Note in
reality, $D$ is an order of magnitude smaller. The advantage of
large $D$ in current simulation is to reduce the Skyrmion radius and
save the calculation resources. The lattice
spacing $a$ is $5\mathring{A}$, and the full simulated sample size is $%
150a\times 50a$, which is much larger than the Skyrmion radius
(about $5a$).
Therefore the finite size effect is safely negligible. Gilbert damping $%
\alpha $ is set to be $0.1$. This value is relatively larger than
the realistic case, but it is helpful to get a relatively larger
stochastic field to make the Skyrmion motions transparent in
simulations.

The phase diagram of the Hamiltonian in Eq.(\ref{eq_discret
Hamiltonian}) is already known\cite{supp,Yu2010a}. The phase
transition between the Skyrmion crystal and the ferromagnetic phase
appears to be of first order and, therefore, the coexistence of both
phases are observed. By tuning the external magnetic field $H$ up to
a critical value, the Skyrmion crystal is melted so that one can
have chance to get a single Skyrmion on the thin
film. The snapshots of a single Skyrmion are shown in Fig.\ref{fig_snapshots}%
(a). Once the magnetic field is further reduced, a perfect Skyrmion
crystal is energetically favored (Fig.\ref{fig_snapshots}(c)). The
color bars in
these plots indicate the topological charge density $q=\frac{1}{4\pi }%
\mathbf{m}\cdot (\partial _{x}\mathbf{m}\times \partial
_{y}\mathbf{m})$. The total topological charge $Q=\int
d^{2}\mathbf{r}q$ counts the number of Skyrmions in the lattice.
%In the case of single Skyrmion, this integral gives unity value.

%%%%%%%%%%%%%%%%%%%%%%%%%%%%%%%%%%%%%%%%%%%%%
\begin{figure}[tbp]
\centering
\includegraphics[width=9cm]{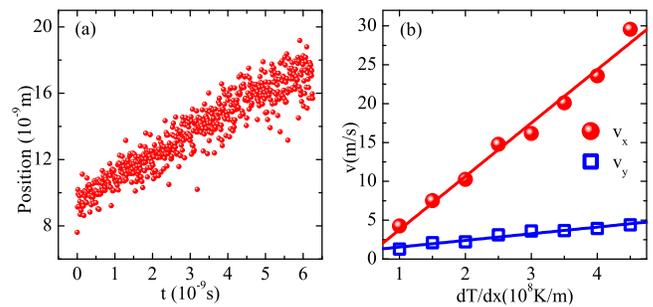}
\caption{(Color online.) (a) A typical simulation showing the
Skyrmion's instant longitudinal positions. Although it fluctuates at
under finite temperature, a forward average velocity is observed.
(b) A linear scaling between the longitudinal velocity and the
temperature gradient (red line). A nonvanishing $v_y$ also addressed
(blue line), indicating the Skyrmion Hall effect.}
\label{fig_velocity}
\end{figure}
%%%%%%%%%%%%%%%%%%%%%%%%%%%%%%%%%%%%%%%%%%%%%%

As the simulation goes on, the single Skymion starts to move under
the effect of the stochastic field. Although the instant velocity
appears to be random, the overall velocity is nonzero.
Quantitatively, we can define the center position $\mathbf{r}_{c}$
of the Skyrmion weighed by the topological
charge: $\mathbf{r}_{c}=\int d^{2}\mathbf{r}\mathbf{m}\cdot (\partial _{x}%
\mathbf{m}\times \partial _{y}\mathbf{m})\mathbf{r}/\int d^{2}\mathbf{r}%
\mathbf{m}\cdot (\partial _{x}\mathbf{m}\times \partial
_{y}\mathbf{m}). $ Fig.\ref{fig_velocity}(a) shows a typical
simulation result of the relation between the center position and
the simulation time. At short time scales, the Skyrmion oscillates
around an average position, in accordance with the thermal
fluctuation. In the long run, the Skyrmion drifts directionally. The
mean velocity is derived by averaging over 1000 simulated events.
Its
relation with the temperature gradient is shown in Fig.\ref{fig_velocity}%
(b). The longitudinal velocity is proportional to the temperature
gradient. Meanwhile, the transverse velocity is nonzero and linear
in temperature gradient as well, although the magnitude is one order
of magnitude smaller than the longitudinal one. The transverse
motion of Skyrmion is another example of the Skyrmion Hall effect in
analogy to the conventional Hall effect for
electrons\cite{Zang2011}.

A surprising result is that the Skyrmion moves from the low
temperature region to the high temperature one, as shown in
Fig.\ref{fig_snapshots}(b). It is generally known that under a
temperature gradient, particles like electrons should move to the
cold terminal, due to the low density of hot particles at the cold
end. This directional Brownian motion gives rise to various
phenomena such as the Seebeck effect. Our result contradicts this
physical picture. This effect even holds also for the entire
Skyrmion crystal in which multiple Skyrmions are driven by the
temperature gradient. As shown in Fig.\ref{fig_snapshots}(d), the
whole crystal shifts towards high temperatures in a similar way.
However, the crystal melts a little bit during the diffusion
process.

%%%%%%%%%%%%%%%%%%%%%%%%%%%%%%%%%%%%%%%
\begin{figure}[tp]
\centering
\includegraphics[width=8cm]{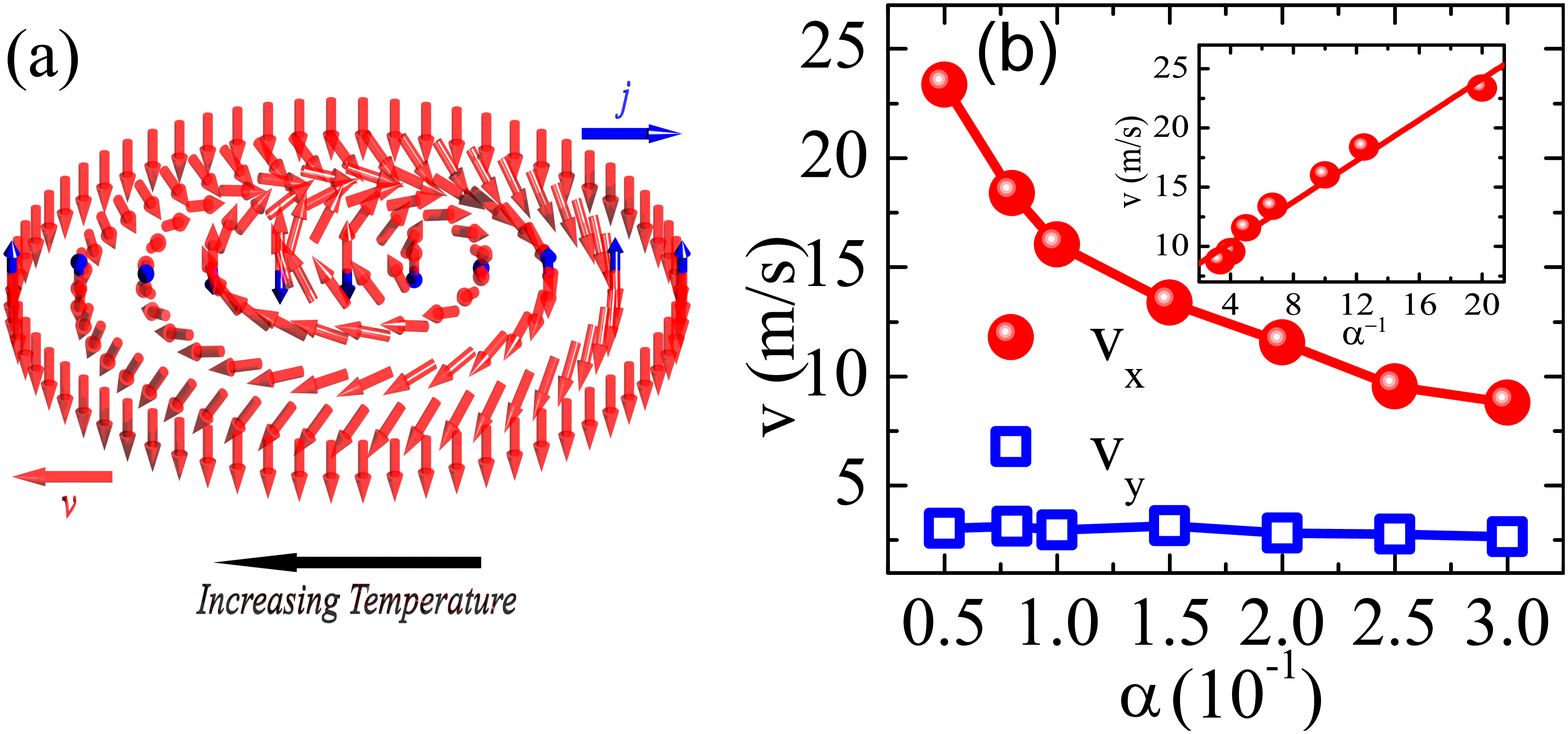}
\caption{(Color online.) (a) Sketch of physical picture for the
Skyrmion motion towards high temperature. Red arrow stands for the
local magnetization constructing the Skyrmion configuration. Blue
arrow stands for the spin of magnon, which points opposite to the
local magnetization. Under a temperature gradient, magnons move from
the high temperature region to the low temperature one indicated by
$j$, resulting in an opposite motion $v$ for the Skyrmion. (b)The
scaling of Skyrmion velocity with the Gilbert damping
$\protect\alpha$. $v_x$ is inversely proportional to
$\protect\alpha$ (red line and the inset). $v_y$ is almost
independent of $\protect\alpha$ (blue line).} \label{fig_sketch}
\end{figure}
%%%%%%%%%%%%%%%%%%%%%%%%%%%%%%%%%%%%%%%%

\textit{Theory:} In order to understand the counter-intuitive
diffusion direction, a magnon assisted theory is
employed\cite{Kovalev2011}. In the simplest case of a ferromagnet
polarized along $\hat{z}$, the spin's deviation $(n_{x},n_{y})$ from
its equilibrium direction is described by the presence of magnons.
The magnon creation operator is $a^{\dagger
}=(n_{x}-in_{y})/\sqrt{2}$, and the magnon number operator is $\rho
=a^{\dagger }a=\frac{1}{2}(n_{x}^{2}+n_{y}^{2})$. The spin component
along
the equilibrium direction is simply $n_{z}=\sqrt{1-(n_{x}^{2}+n_{y}^{2})}%
\approx 1-\rho $. This result shows that each magnon carries spin
one polarized antiparallelly with the equilibrium direction.
Therefore, once
there is a Skyrmion under a temperature gradient, as shown in Fig.\ref%
{fig_sketch}(a), magnon as a low-lying excitation responses much
more actively than the Skyrmion itself. As a typical quasiparticle,
the magnon diffuses from hot to cold end in the usual way. Due to
the antiparallel alignment of the spin, the magnon current provides
a negative transfer torque on the Skyrmion. Due to the conservation
of the total angular momentum, the Skyrmion moves in the opposite
way.

To quantitatively formulate this physical picture, let's decompose
the local magnetizations into the slow mode $\mathbf{m}_{s}$ and the
orthogonally fast
mode $\mathbf{m}_{f}=\mathbf{m}_{s}\times \mathbf{n}$: $\mathbf{m}=(1-%
\mathbf{m}_{f}^{2})^{1/2}\mathbf{m}_{s}+\mathbf{m}_{f}$. The slow
mode is responsible for the equilibrium configuration of the
Skyrmion. Substituting
it into the continuum version of the Hamiltonian $H=\int d^{2}r[\frac{1}{2}%
J(\nabla \mathbf{m})^{2}+\frac{D}{a}\mathbf{m}\cdot (\nabla \times \mathbf{m}%
)-\frac{\mathbf{H}}{a^{2}}\cdot \mathbf{m}]$ and keeping only the
dominant terms arising from the fluctuations of fast mode, one can
get the following equation of motion for the slow
mode\cite{supp,Kovalev2011}:
\begin{equation}
\mathbf{\dot{m}}_{s}=-\gamma Ja^{2}\mathbf{j}\cdot \bm{\nabla}\mathbf{m}%
_{s}-\gamma \mathbf{m}_{s}\times \mathbf{L}+\alpha
\mathbf{m}_{s}\times \mathbf{\dot{m}}_{s}  \label{eq_current driven}
\end{equation}%
where $j_{i}=\mathbf{m}_{s}\cdot (\mathbf{n}\times \partial _{i}\mathbf{n}%
)=i(\partial _{i}a^{\dagger }a-a^{\dagger }\partial _{i}a)$ is the
magnon current induced by the temperature gradient. Note that the
first term on the right hand side of Eq.(\ref{eq_current driven}) is
analogous to the spin
transfer torque provided by the itinerant electrons in the adiabatic limit%
\cite{Zang2011,Tatara2008}. However the sign is different in the two
cases. The negative sign here corresponds to the negative transfer
torque from the magnons.

Ignoring the deformation of the Skyrmion, the slow modes can be
written in
terms of the collective coordinates $\mathbf{u}(t)$ as $\mathbf{m}_{s}(r,t)=%
\mathbf{m}_{s}^{0}(\mathbf{r}-\mathbf{u}(t))$, where
$\mathbf{m}_{s}^{0}$ is the ground configuration, and
$\mathbf{u}(t)$ describes the position of the Skyrmion. Inserting it
into Eq.(\ref{eq_current driven}) and integrating over the ground
configuration, one finally gets the equation motion for the
collective coodinates $Q\varepsilon ^{ij}\dot{u}_{j}(t)=Q\gamma
Ja^{2}\varepsilon ^{ij}j_{j}+2\alpha \eta
\dot{u}_{i}(t)+\frac{\gamma }{4\pi }\int d^{2}r\partial
_{i}\mathbf{m}_{s}^{0}\cdot \mathbf{L}(\mathbf{r+u},t),$
where the shape factor $\eta =\frac{1}{8\pi }\int d^{2}r\partial _{i}\mathbf{%
m}_{s}^{0}\times \partial _{i}\mathbf{m}_{s}^{0}$ is close to unity.
We define a collective stochastic force $l_{i}$ acting on the
Skyrmion as a
whole by $l_{i}(\mathbf{u},t)=\int d^{2}r\partial _{i}\mathbf{m}%
_{s}^{0}\cdot \mathbf{L(r+u,t)}$, whose average then satisfies
$\left\langle l_{i}(\mathbf{u},t)l_{j}(\mathbf{u}^{\prime
},t^{\prime })\right\rangle =\xi ^{\prime }\delta _{ij}\delta
(\mathbf{u-u}^{\prime })\delta (t-t^{\prime })$ with a new mean
square $\xi ^{\prime }=8\pi \eta \xi =8\pi \alpha \eta
a^{2}k_{B}T/\gamma $. The collective equation of motion resembles
the standard Langevin equation. Let $P(\mathbf{r},t)$ be the
probability to find the Skyrmion at position $\mathbf{r}$ and the
time $t$. It thus satisfies
the Fokker-Planck equation\cite{Fokker}: $\frac{\partial P}{\partial t}%
=-[\gamma Ja^{2}j_{x}-2(\frac{\gamma }{4\pi Q})^{2}(\partial _{x}\xi
^{\prime })]\partial _{x}P-\gamma Ja^{2}j_{x}\cdot 2\alpha \eta
\partial _{y}P+(\frac{\gamma }{4\pi Q})^{2}\xi ^{\prime }(\partial
_{x}^{2}+\partial _{y}^{2})P$. At the current stage, we are only
interested in the lowest
order traveling wave solution of the Fokker-Planck equation, namely $P(%
\mathbf{r},t)=P(\mathbf{r}-\mathbf{v}t)$. The last term provides
nonlinearity: it thus broadens the wave package and can be
neglected. Finally, we get the average velocity of the Skyrmion in
both the longitudinal and transverse directions
\begin{eqnarray}
v_{x} &=&\gamma Ja^{2}j_{x}-\frac{\gamma }{\pi Q^{2}}\alpha \eta a^{2}k_{B}%
\frac{dT}{dx}\equiv v_{x}^{M}-v^{B}  \label{eq_vx} \\
v_{y} &=&2\alpha \eta v_{x}^{M}
\end{eqnarray}%
The contributions from the magnon and the Brownian motion are
separable and are denoted, respectively, by $v_{x,y}^{M}$ and
$v^{B}$. Eq.(\ref{eq_vx}) shows explicitly that their effects are
completely opposite: the Skyrmion is pushed by the Brownian motion
towards the cold terminal, while it is pulled back to the hot end by
the magnon. On the other hand, as the temperature
gradient is exerted along the $x$ direction, the Brownian motion along the $%
y $ direction vanishes on the average. Only the magnon effect
contributes to the transverse velocity, which is a factor $\alpha $
smaller than the
longitudinal one, agreeing with the numerical simulation in Fig.\ref%
{fig_velocity}b. This Hall effect of the Skyrmion motion is closely
related to the topology of the Skyrmion texture captured by the
nonzero topological charge $Q$. Generally speaking, a directional
transverse motion requires the
breaking of time reversal symmetry. Here it is the dissipative damping $%
\alpha $ that breaks time reversal. Therefore one expects the
proportionality between transverse velocity and Gilbert damping
$\alpha $.

Now it is important to evaluate the magnon current. To this end, we
can
apply a semi-classical approach with the relaxation time approximation\cite%
{Ashcroft}. The variation from the Bose-Einstein distribution $f$ is
given by $\delta f=\tau \frac{\partial f}{\partial \varepsilon
}\frac{\varepsilon }{T}\mathbf{v}\cdot \nabla T$. $\tau $ is the
relaxation time. The magnon
current is consequently $j=a^{2}\int \frac{d^{2}k}{(2\pi )^{2}}%
k_{x}a_{k}^{\dagger }a_{k}=a^{2}\int \frac{d^{2}k}{(2\pi )^{2}}\tau k_{x}%
\frac{\partial f}{\partial \varepsilon }\frac{\varepsilon
}{T}\frac{\partial \varepsilon }{\hbar \partial
k_{x}}\frac{dT}{dx}$. In this simple evaluation, higher order
processes such as magnon-magnon interactions are neglected so that
$\tau $ is given by the Gilbert damping only. In the presence of a
nonzero $\alpha $, the magnon frequency acquires an imaginary
value $\alpha \omega $. Therefore the magnon number decays exponentially as $%
\rho (t)\sim \exp (-2\alpha \omega t)$. The relaxation time is thus
$\tau
=1/(2\alpha \omega )$. According to the work by Petrova and Tchernyshyov\cite%
{Petrova2011}, linear dispersion is respected in the Skyrmion
crystal, given by $\varepsilon \hbar \omega
=\frac{1}{2}M_{0}Da\gamma \hbar k\equiv s\hbar k $, where $M_{0}$ is
the magnitude of the local spin. $s$ is the effective velocity of
the magnon. Finally one gets the magnon current given by
\begin{equation}
j=j_{x}=\frac{\pi }{24}a^{2}(\frac{k_{B}}{\hbar s})^{2}\frac{T}{\alpha }%
\frac{dT}{dx}  \label{eq_magnon current}
\end{equation}%
This result indicates that the magnon current, as well as the
Skyrmion velocity, is proportional to the temperature gradient. It
is quite consistent with the numerical result in
Fig.\ref{fig_velocity}. This evaluation also explains the reason why
the magnon contribution is overwhelming in Eq.(\ref{eq_vx}). The
ratio between these two contributions is
$v^{B}/v_{x}^{M}=\frac{6}{\pi }\alpha ^{2}\frac{D^{2}}{Jk_{B}T}$. It
is definitely a small number due to the small DM interaction and the
tiny damping coefficient. The net effect of the Brownian motion is
almost invisible in this case.

Another interesting conclusion from Eq.(\ref{eq_magnon current}) is
that $j$ is inversely proportional to $\alpha$. Consequently, the
longitudinal Skyrmion velocity is also inversely proportional to
$\alpha $, while the transverse velocity is independent of $\alpha
$. In reality, dislocations or imperfections of the Skyrmion lattice
may affect the magnon dispersion significantly.
%The magnon dispersion of a single Skyrmion would be distorted in a complicated way.
However, as long as the magnon-magnon interaction is negligible, the
inverse proportionality between the longitudinal velocity and the
Gilbert damping always holds. In the case of insulating helimagnets,
as the magnetization energy can hardly be dissipated away, the
Gilbert damping is tiny. The Skyrmion velocity can be quite large
instead. In order to test this theory, we scaled the velocity with
respect to $\alpha $ from the simulations, as
shown in Fig.\ref{fig_sketch}(b). A nice inverse proportionality between $%
v_{x}$ and $\alpha $ is explicitly addressed. $v_{y}$ remains almost
the same for different $\alpha $ values. These results match well
with our theory.

A similar magnon assisted theory was applied to the case of domain
wall motion\cite{Hinzke,Kovalev2011}. However, the difference
brought by the topology of the Skyrmion is profound. In the
derivation of collective equation of motion, the quantization of
topological charge $Q$ is applied, which is the key feature of the
Skyrmion. For the domain wall case, the total topological charge
vanishes, so that this method doesn't apply. The collective equation
of motion for the Skyrmion provides us a universal dynamics that
weakly depends on the detailed structure of the Skyrmion.
Furthermore, the Skyrmion stability allows us to treat it as a
quasiparticle, so that the Fokker-Planck equation comes into play.
The generalization from 1D domain wall to 2D Skyrmion crystal brings
new phenomena such as the Skyrmion Hall effect.

\textit{Estimates:} For Cu$_{2}$OSeO$_{3}$, $J/k_{B}\sim 50$K,
$M_{0}=1/2$,
and the spiral period is $\lambda \approx 2\pi Ja/D\sim 50$nm\cite{Seki2012}%
. Let $a\sim 5\mathring{A}$, thus $D/k_{B}\sim 3$K, and the
effective velocity $s\sim 15.6$m/s. As a reasonable estimate, let
$\alpha =0.01$, then our theory gives $v_{x}\sim 1.2\times
10^{-4}\frac{dT}{dx}(m/s)$. In the numerical simulations, $v_{x}\sim
10^{-7}\frac{dT}{dx}$ for $\alpha =0.1$. This three orders of
magnitude of difference can be perfectly fixed by noting a
difference of a factor $\sim $ $10$ in DM interaction $D$, and
another factor $10$ in the Gilbert damping $\alpha $. This serves as
a quantitative confirmation of our theory. Experimentally a
reasonably large temperature gradient is about $1$K per
millimeter\cite{Uchida2008}, so that a velocity of $0.1$m/s can be
achieved. This Skyrmion motion can be traced by real-space
spectroscopies such as the Lorentz TEM. Another interesting issue is
how to observe this phenomenon by transport measurements. As
itinerant electrons are absent, signals of the topological Hall
effect present in metallic Skyrmion crystals\cite{Neubauer2009,
Huang2012} is missing here. Probably the measurement technique of
the spin Seebeck effect would come to help as a moving Skyrmion
carries a spin current. There's no spin polarized electron current
in this system, so that the signal of the moving Skyrmion would be
dominant.

Compared to the domain wall motion, a peculiar advantage of the
Skyrmion motion is its tiny pinning indicated by the small threshold
current in current driven case. This pinning results from impurities
and lattice imperfections, which have basically the same level in
insulating and metallic Skyrmion crystals. A low critical current of
$10^{6}$A/m$^{2}$ is observed in MnSi\cite{Jonietz2010}, which
corresponds to a theoretical velocity of
$10^{-4}$m/s\cite{Zang2011}. Our estimate of the temperature
gradient driven Skyrmion motion is far beyond this threshold, thus
it can be easily realized. The interaction between magnon and
Skyrmion discussed here might open a new field of 'Skyrmionic
magnonics'.

We are grateful for the insightful discussions with A. Abanov,
Cosimo Bambi, S. X. Huang, N. Nagaosa, V. Vakaryuk, O. Tchernyshyov,
Y. Tserkovnyak, and Hui Wang.This work was supported in part by the
Theoretical Interdisciplinary Physics and Astrophysics Center and by
the U.S. Department of Energy, Office of Basic Energy Sciences,
Division of Materials Sciences and Engineering under Award
DEFG02-08ER46544, and Fudan Research Program on Postgraduates.

\section{Numerical Details}

The lattice Hamiltonian in the numerical simulations is given by%
\begin{eqnarray}
H &=&-J\sum_{\mathbf{r}}\mathbf{m}_{\mathbf{r}}\cdot (\mathbf{m}_{\mathbf{r}+%
\hat{x}}+\mathbf{m}_{\mathbf{r}+\hat{y}})  \notag \\
&&-D\sum_{\mathbf{r}}[\hat{x}\cdot (\mathbf{m}_{\mathbf{r}}\times \mathbf{m}%
_{\mathbf{r}+\hat{x}})+\hat{y}\cdot (\mathbf{m}_{\mathbf{r}}\times \mathbf{m}%
_{\mathbf{r}+\hat{y}})]  \notag \\
&&-H\sum_{\mathbf{r}}\mathbf{m}_{\mathbf{r}}^{z}
\end{eqnarray}%
where the first term is the ferromagnetic Heisenberg exchange, the
second term is the Dzyaloshinskii-Moriya interaction, and the last
term is the
Zeeman coupling. The effective field $\mathbf{H}_{eff}$ acting on $\mathbf{m}%
_{r}$ is therefore given by%
\begin{eqnarray}
\mathbf{H}_{eff} &=&-\partial H/\partial \mathbf{m}_{r} \\
&=&J(\mathbf{m}_{\mathbf{r}+\hat{x}}+\mathbf{m}_{\mathbf{r}+\hat{y}})+D(%
\mathbf{m}_{\mathbf{r}+\hat{x}}\times \hat{x}+\mathbf{m}_{\mathbf{r}+\hat{y}%
}\times \hat{y})+H\hat{z}  \notag
\end{eqnarray}%
In the absence of external magnetic field $H$, a helical order is
favored at low temperature. The reason is the following. As
Dzyaloshinskii-Moriya interaction breaks inversion symmetry, its
Fourier transformation gives a term linear in wavevector $k$,
contrary to the $k^{2}$ term contributed by
the Heisenberg exchange. Energy can be minimized at a nonzero wavevector $%
k\sim D/J$, and the helical state is generated. However, the total
magnetization in the helical state is zero, so that it cannot save
energy from the Zeeman term. Once magnetic field is turned on, a
phase transition from the helical state to the Skyrmion crystal
state is observed. The large magnetic field limit corresponds to a
ferromagnetic state.
%%%%%%%%%%%%%%%%%%%%%%%%%%%%%%%%%
\begin{figure}[tbp]
\centering
\includegraphics[width=9cm]{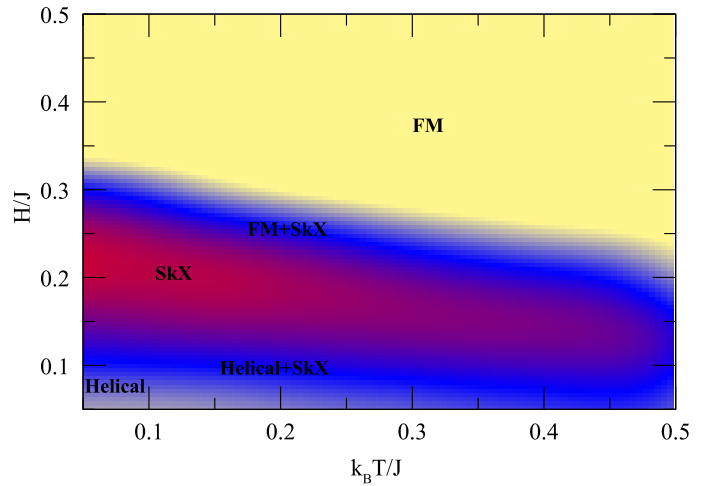}
\caption{The phase diagram of the model Hamiltonian.} \label{fig_S1}
\end{figure}
%%%%%%%%%%%%%%%%%%%%%%%%%%%%%%%%%

In our simulations, $D=0.5J$. The finite temperature phase diagram
is shown in Fig. 1. It shows explicitly that the Skyrmion crystal
phase is stabilized down to zero temperature. The phase transition
from the Skyrmion crystal to ferromagnetic state is believed to be
first order, so that dilute Skyrmions are observed at a narrow
magnetic field/temperature window. In the dynamical simulations, the
Skyrmion crystal corresponds to $H=0.15J$, while a single Skyrmion
is generated at $H=0.3J$.

In the stochastic LLG equation, the choice of time step for
numerical integration is an important issue. In each simulated
event, the final time is set to be $1000\hbar/J=9.6\times10^{-10}$s.
The mean velocity is derived by averaging over $1000$ simulated
events. Fig. 2 shows the relation between average longitudinal
velocity and the time step. It shows the result is convergent once
the time step is smaller than $0.06$ps. In the simulations in the
main content, a time step of $0.05\hbar/J=4.8\times10^{-14}$s is
employed.

\section{Derivation of Eq. (5)}

In order to derive an equation of motion describing the effect of
magnons on the equilibrium magnetization configurations, one has to
separate these two
degrees of freedom. To this end, let's decompose each local magnetization $%
\mathbf{m}$ into slow modes $\mathbf{m}_{s}$ and orthogonally fast modes $%
\mathbf{m}_{f}$
\begin{eqnarray}
\mathbf{m}
&=&(1-\mathbf{m}_{f}^{2})^{1/2}\mathbf{m}_{s}+\mathbf{m}_{f}
\notag \\
&\approx &\mathbf{m}_{s}+\mathbf{m}_{f}-\frac{1}{2}(\mathbf{m}_{f}^{2})%
\mathbf{m}_{s}+O(\mathbf{m}_{f}^{3})
\end{eqnarray}%
%
%
%
%
%
%
%
%%%%%%%%%%%%%%%%%%%%%%%%%%%%%%%%%
\begin{figure}[tbp]
\centering
\includegraphics[width=9cm]{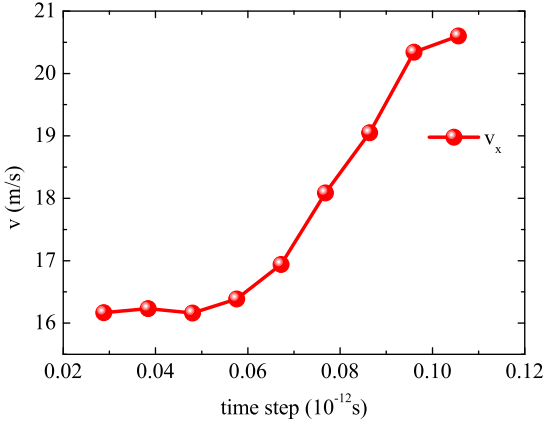}
\caption{The relation between the average longitudinal velocity and
the time step in the numerical integration. Convergence is achieved
once the time step is smaller than $0.06$ps.} \label{fig_S2}
\end{figure}
%%%%%%%%%%%%%%%%%%%%%%%%%%%%%%%%%
where $\mathbf{m}_{s}$ describes the equilibrium configuration, while $%
\mathbf{m}_{f}$ is responsible for the magnonic dynamics. As the
Hamiltonian is given by%
\begin{equation}
H=\int d^{2}r[\frac{1}{2}J(\nabla
\mathbf{m})^{2}+\frac{D}{a}\mathbf{m}\cdot (\nabla \times
\mathbf{m})-\frac{\mathbf{H}}{a^{2}}\cdot \mathbf{m}],
\end{equation}%
the effective field $\mathbf{H}_{eff}$ in the Stochastic
Landau-Lifshitz-Gilbert equation $\mathbf{\dot{m}}=-\gamma \mathbf{m}\times (%
\mathbf{H}_{eff}+\mathbf{L})+\alpha \mathbf{m}\times
\mathbf{\dot{m}}$ is
therefore%
\begin{eqnarray}
\mathbf{H}_{eff} &=&-\partial H/\partial \mathbf{m}  \notag \\
&=&Ja^{2}\nabla ^{2}\mathbf{m}-2Da\nabla \times \mathbf{m+H} \\
&=&Ja^{2}\nabla _{a}[\nabla _{a}\mathbf{m}_{s}-\mathbf{m}_{f}(\nabla _{a}%
\mathbf{m}_{f})\mathbf{m}_{s}-\frac{1}{2}\mathbf{m}_{f}^{2}\nabla _{a}%
\mathbf{m}_{s}  \notag \\
&&+\nabla _{a}\mathbf{m}_{f}]-2Da[\nabla \times
\mathbf{m}_{s}+\nabla \times
\mathbf{m}_{f}  \notag \\
&&-\mathbf{m}_{f}(\nabla \mathbf{m}_{f})\times \mathbf{m}_{s}-\frac{1}{2}%
\mathbf{m}_{f}^{2}\nabla \times \mathbf{m}_{s}]+\mathbf{H}+o(\mathbf{m}%
_{f}^{3})  \notag
\end{eqnarray}%
As $\mathbf{m}_{f}$ has a characteristic length scale of lattice constant $a$%
, compared to a length scale of $(J/D)a$ for $\mathbf{m}_{s}$, it
contributes larger derivatives. Therefore the leading terms of the
torque are listed in the following
\begin{equation*}
-\mathbf{m}\times \mathbf{H}_{eff}\approx
-Ja^{2}\mathbf{m}_{s}\times \nabla
^{2}\mathbf{m}_{f}-Ja^{2}\mathbf{m}_{f}\times \nabla
^{2}\mathbf{m}_{f}
\end{equation*}%
Here the DM terms on $\mathbf{m}_{f}$ and $\mathbf{m}_{s}$ can be
completely neglected as only one spatial derivative will be taken
into account, which is one order of magnitude smaller than the
Heisenberg term. Contributions
from the Zeeman term are negligible in a similar way. Besides, as $\mathbf{m}%
_{f}$ is a fast mode behaving as a function of
sine or cosine in time, by taking the average over time, linear terms in $%
\mathbf{m}_{f}$ can be discarded as well. Therefore%
\begin{eqnarray}
-\mathbf{m}\times \mathbf{H}_{eff} &\approx
&-Ja^{2}\mathbf{m}_{f}\times
\nabla ^{2}\mathbf{m}_{f} \\
&=&-Ja^{2}\nabla (\mathbf{m}_{f}\times \nabla \mathbf{m}_{f})
\end{eqnarray}%
Noticing $\mathbf{m}_{f}\perp \mathbf{m}_{s}$, let $\mathbf{m}_{f}=\mathbf{m}%
_{s}\times \mathbf{n}$, where $\mathbf{n}$ is an arbitrary vector
perpendicular to $\mathbf{m}_{s}$. Substitute it into the expression
of the torque, then
\begin{eqnarray}
\mathbf{m}_{f}\times \nabla \mathbf{m}_{f} &=&(\mathbf{m}_{s}\times \mathbf{n%
})\times \nabla (\mathbf{m}_{s}\times \mathbf{n})  \notag \\
&=&(\mathbf{m}_{s}\times \mathbf{n})\times (\nabla
\mathbf{m}_{s}\times
\mathbf{n+m}_{s}\times \nabla \mathbf{n})  \notag \\
&=&-\mathbf{n}\nabla \mathbf{m}_{s}\cdot (\mathbf{m}_{s}\times \mathbf{n})+%
\mathbf{m}_{s}\nabla \mathbf{n\cdot }(\mathbf{m}_{s}\times
\mathbf{n})
\notag \\
&\approx &\mathbf{m}_{s}\nabla \mathbf{n\cdot }(\mathbf{m}_{s}\times \mathbf{%
n})  \notag \\
&=&\mathbf{m}_{s}\mathbf{j}
\end{eqnarray}%
where $\mathbf{j}=\nabla \mathbf{n\cdot }(\mathbf{m}_{s}\times
\mathbf{n})$ is the magnon current defined in the main content. Here
again we applied the
fact that $\nabla \mathbf{n}$ is overwhelming. Consequently%
\begin{equation}
\nabla (\mathbf{m}_{f}\times \nabla \mathbf{m}_{f})=(\partial _{\mu }\mathbf{%
m}_{s})j_{\mu }+\mathbf{m}_{s}\partial _{\mu }j_{\mu }
\end{equation}%
Assume magnon current is steady so that $\nabla\cdot\mathbf{j}=0$,
then the torque is given by
\begin{equation}
-\mathbf{m}\times \mathbf{H}_{eff}=-Ja^{2}j_{\mu }(\partial _{\mu }\mathbf{m}%
_{s})
\end{equation}%
The stochastic part of the spin transfer torque is s\bigskip imply%
\begin{equation}
-\gamma \mathbf{m}\times \mathbf{L=}-\gamma (\mathbf{m}_{s}+\mathbf{m}%
_{f})\times \mathbf{L}\approx -\gamma \mathbf{m}_{s}\times
\mathbf{L}
\end{equation}%
The term $-\gamma \mathbf{m}_{f}\times \mathbf{L}$ is neglected as
it is a linear function of $\mathbf{m}_{f}$ only, whose time average
vanishes. Due to the same reason, the Gilbert damping term is
\begin{equation}
\alpha \mathbf{m}\times \mathbf{\dot{m}}\approx \alpha
\mathbf{m}_{s}\times \mathbf{\dot{m}}_{s}+\alpha
\mathbf{m}_{f}\times \mathbf{\dot{m}}_{f}
\end{equation}%
As discussed above, $\mathbf{m}_{f}$ is completing cyclotron
rotations with respect to $\mathbf{m}_{s}$, which in basically a
sine or cosine function in time. The time derivative turns a sine
function to cosine and vice versa.
Therefore the time average of $\alpha \mathbf{m}_{f}\times \mathbf{\dot{m}}%
_{f}$ vanishes so that
\begin{equation}
\alpha \mathbf{m}\times \mathbf{\dot{m}}\approx \alpha
\mathbf{m}_{s}\times \mathbf{\dot{m}}_{s}
\end{equation}%
As a result, the LLG equation for $\mathbf{m}_{s}$ is given by%
\begin{equation}\label{eq_EOM_magnon}
\mathbf{\dot{m}}_{s}=-\gamma Ja^{2}j_{\mu }\partial _{\mu }\mathbf{m}%
_{s}-\gamma \mathbf{m}_{s}\times \mathbf{L}+\alpha
\mathbf{m}_{s}\times \mathbf{\dot{m}}_{s}
\end{equation}%
which is exactly the Eq. (5) in the main content.

The magnon part of this equation of motion shows similarity with the
magnetization dynamics in the presence of a charge current, which is
given by\cite{Bazaliy,Tatara}%
\begin{equation}\label{eq_EOM_metal}
\mathbf{\dot{m}}_{s}=\frac{\hbar \gamma }{2e}j_{\mu }\partial _{\mu }\mathbf{%
m}_{s}
\end{equation}%
here $\mathbf{j}$ is the electric current flowing through the
magnetic sample. In deriving this equation, adiabatic approximation
is applied such that the spins of itinerant electrons are always
parallel with the local magnetizations $\mathbf{m}_{s}$. The steady
charge current contributes a term of $H=-\frac{1}{c}\int d^{2}x\mathbf{j}%
\cdot \mathbf{a}$ to the spin Hamiltonian and leads to the equation
of
motion Eq. (\ref{eq_EOM_metal}). Here $\mathbf{a}_{\mu }=\frac{\hbar c}{2e}%
(1-\cos \theta )\partial _{\mu }\varphi $ is the emergent gauge
field
associated with the magnetization configuration $\mathbf{m}_{s}$, where $%
\theta $ and $\varphi $ are the inclination and azimuthal angles of $\mathbf{%
m}_{s}$ respectively\cite{Zang}. One can easily show that the
corresponding emergent
magnetic field $\mathbf{b}=\nabla \times \mathbf{a=}\frac{\hbar c}{2e}%
\mathbf{m}_{s}\cdot (\partial _{x}\mathbf{m}_{s}\times \partial _{y}\mathbf{m%
}_{s})$ counts the topological charge density.

The similarity between the magnon driven magnetization dynamics and
the current driven one indicates that the magnon current
$\mathbf{j}$ couples to the emergent gauge field $\mathbf{a}$ in the
same way. The reason comes from the key physical picture introduced
in the main content. The magnon naturally obeys the adiabatic
approximation in the leading order as the spin
of a magnon is nothing but the reverse of its equilibrium direction $\mathbf{%
m}_{s}$. The validity of adiabaticity leads to the same structure as
the current driven one. The only difference is the sign in front the
$j_{\mu }\partial _{\mu }\mathbf{m}_{s}$ in Eq.
(\ref{eq_EOM_magnon}) and Eq. (\ref{eq_EOM_metal}). This is because
the magnetization $\mathbf{m}_{s}$ is parallel with the spin of
itinerant electron, while anti-parallel with the spin of magnon.


\begin{thebibliography}{99}
\bibitem{skyrme1961} T. H. R. Skyrme, Proc. Roy. Soc. London A \textbf{260},
127 (1961); Nuc. Phys. \textbf{31}, 556 (1962).

\bibitem{Bogdanov} A.N. Bogdanov and D. A. Yablonskii, Sov. Phys. JETP
\textbf{68}, 101-103 (1989)

\bibitem{Rosler} U.K. R\"{o}{\ss }ler, A.N. Bogdanov, and C. Pfleiderer,
Nature 4\textbf{42}, 797-801 (2006).

\bibitem{Muehlbauer2009a} S. M\"{u}hlbauer, B. Binz, F. Joinetz, C.
Pfleiderer, A. Rosch, A. Neubauer, R. Georgii, and P. B\"{o}ni,
Science \textbf{323}, 915 (2009).

\bibitem{Yu2010a} X. Z. Yu \textit{et al.}, Nature (London) \textbf{465},
901 (2010).

\bibitem{Yi2009} S. D. Yi, S. Onoda, N. Nagaosa, and J. H. Han, Phys. Rev. B
\textbf{80}, 054416 (2009).

\bibitem{Rosler2010} U. K. R{\"o\ss }ler, A. A. Leonov, and A. N. Bogdanov,
arXiv:1009.4849.

\bibitem{Han2010} J. H. Han, J. Zang, Z. Yang, J. H. Park, and N. Nagaosa,
Phys. Rev. B \textbf{82}, 094429 (2010).

\bibitem{Seki2012} S. Seki, X. Z. Yu, S. Ishiwata, and Y. Tokura, Science,
\textbf{336}, 198 (2012).

%\bibitem{Liu2012} Y. H. Liu, Y. Q. Li, and J. H. Han, arXiv: 1209.3120.

\bibitem{Yu2012} X. Z. Yu \textit{et al.}, PNAS \textbf{109}, 8856 (2012).

\bibitem{Zang2011} J. Zang, M. Mostovoy, J. H. Han, and N. Nagaosa, Phys.
Rev. Lett. \textbf{107}, 136804 (2011).

\bibitem{Schulz2012} T. Schulz \textit{et al.}, Nat. Phys. \textbf{8}, 301
(2012).

\bibitem{Mochizuki2012} M. Mochizuki, Phys. Rev. Lett. \textbf{108}, 017601
(2012).

\bibitem{Jonietz2010} F. Jonietz \textit{et al.}, Science \textbf{330}, 1648
(2010).

\bibitem{Evershor2012} K. Everschor, M. Garst, B. Binz, F. Jonietz, S. M\"{u}%
hlbauer, C. Pfleiderer, and A. Rosch, Phys. Rev. B \textbf{86},
054432 (2012).

\bibitem{Palacios1998} J. L. Garc\'{\i}a-Palacios, and F. J. L\'{a}zaro,
Phys. Rev. B \textbf{58}, 14937 (1998).

\bibitem{Hinzke} D. Hinzke and U. Nowak, Phys. Rev. Lett. \textbf{107},
027205 (2011).

\bibitem{supp} See supplemental material.

\bibitem{Kovalev2011} A. A. Kovalev, and Y. Tserkovnyak, Europhys. Lett.
\textbf{97}, 67002 (2012).

\bibitem{Tatara2008} G. Tatara, H. Kohno, and J. Shibata, Phys. Rep. \textbf{%
468}, 213 (2008).

\bibitem{Fokker} H. Risken, \textit{The Fokker-Planck Equation}, 2nd ed.
(Springer, Berlin, 1989).

\bibitem{Ashcroft} N. W. Ashcroft, and N. D. Mermin, \textit{Solid State
Physics}, Saunders College 1976.

\bibitem{Petrova2011} O. Petrova, and O. Tchernyshyov, Phys. Rev. B \textbf{%
84}, 214433 (2011).

\bibitem{Uchida2008} K. Uchida, S. Takahashi, K. Harii, J. Ieda, W.
Koshibae, K. Ando, S. Maekawa and E. Saitoh, Nature \textbf{45}, 778
(2008).

\bibitem{Neubauer2009} A. Neubauer \textit{et al.}, Phys. Rev. Lett. \textbf{%
102}, 186602 (2009).

\bibitem{Huang2012} S. X. Huang, and C. L. Chien, Phys. Rev. Lett. \textbf{%
108}, 267201 (2012).
\end{thebibliography}

\begin{thebibliography}{99}
\bibitem{Bazaliy} Ya.B. Bazaliy, B.A. Jones, and S. C. Zhang, Phys. Rev. B
\textbf{57}, R3213 (1998).

\bibitem{Tatara} G. Tatara, H. Kohno, and J. Shibata, Phys. Rep. \textbf{468}, 213
(2008).

\bibitem{Zang} J. Zang, M. Mostovoy, J. H. Han, and N. Nagaosa,
Phys. Rev. Lett. \textbf{107}, 136804 (2011)

\end{thebibliography}
\end{document}